\documentclass[doublespacing]{elsart}

\usepackage{natbib}

\usepackage{longtable}
\usepackage{lscape}
\usepackage{setspace}
\usepackage{lineno}
  
\bibpunct{(}{)}{;}{a}{,}{,}

\usepackage{graphicx}

\usepackage{amssymb}
\usepackage{wasysym}

\newcommand{\BibTeX}{ \textrm{B\kern-.05em\textsc{i\kern-.025em b}\kern-.08em
    T\kern-.1667em\lower.7ex\hbox{E}\kern-.125emX} }

\begin{document}

\begin{frontmatter}



\title{Near-infrared thermal emission from near-Earth asteroids: Aspect-dependent variability}


\author[lowell,mit]{Nicholas A. Moskovitz\thanksref{label}}
\author[wi]{David Polishook}
\author[mit]{Francesca E. DeMeo}
\author[mit]{Richard P. Binzel}
\author[umass]{Thomas Endicott}
\author[uh,eso]{Bin Yang}
\author[lpl]{Ellen S. Howell}
\author[jhu]{Ronald J. Vervack, Jr.}
\author[ucf]{Yanga R. Fern\'andez}

\address[lowell]{Lowell Observatory, 1400 West Mars Hill Road, Flagstaff, AZ 86001 (U.S.A.)}

\address[wi]{Department of Earth and Planetary Sciences, Weizmann Institute of Science, Rehovot 7610001 (Israel)}

\address[mit]{Massachusetts Institute of Technology, Department of Earth, Atmospheric and Planetary Sciences, 77 Massachusetts Avenue, Cambridge, MA 02139 (U.S.A.)}

\address[umass]{University of Massachusetts, Boston, 100 Morrissey Blvd., Boston, MA 02125 (U.S.A.)}

\address[uh]{Institute for Astronomy, University of Hawaii, 2680 Woodlawn Drive, Honolulu, HI 96822 (U.S.A.)}

\address[eso]{European Southern Observatory, Alonso de Cordova, 3107, Vitacura, Casilla 19001, Santiago de Chile (Chile)}

\address[lpl]{Lunar and Planetary Lab, University of Arizona, Tucson AZ 85721}

\address[jhu]{JHU/Applied Physics Lab, 11100 Johns Hopkins Road, Laurel MD 20723 (U.S.A.)}

\address[ucf]{University of Central Florida, Dept. of Physics, 4000 Central Florida Blvd. Orlando FL 32828 (U.S.A.)}

 \thanks[label]{Observations conducted while at MIT, current affiliation is Lowell Observatory.}

\begin{center}
\scriptsize
Copyright \copyright\ 2016 Nicholas A. Moskovitz
\end{center}


%
%
%
%
%


\end{frontmatter}



\begin{flushleft}
\vspace{1cm}
Number of pages: \pageref{lastpage} \\
Number of tables: \ref{lasttable}\\
Number of figures: \ref{lastfig}\\
\end{flushleft}



\clearpage
{\bf Proposed Running Head:}
NEA Thermal Emission
\\
\\
{\bf Please send Editorial Correspondence to:}
\\
Nicholas A. Moskovitz\\
Lowell Observatory\\
1400 West Mars Hill Road\\
Flagstaff, AZ 86001, USA. \\
\\
Email: nmosko@lowell.edu\\
Phone: (928) 779-5468 \\

\clearpage


\begin{abstract}

Here we explore a technique for constraining physical properties of near-Earth asteroids (NEAs) based on variability in thermal emission as a function of viewing aspect. We present case studies of the low albedo, near-Earth asteroids (285263) 1998 QE2 and (175706) 1996 FG3. The Near-Earth Asteroid Thermal Model (NEATM) is used to fit signatures of thermal emission in near-infrared (0.8 - 2.5 $\mu$m) spectral data. This analysis represents a systematic study of thermal variability in the near-IR {\it as a function of phase angle}. The observations of QE2 imply that carefully timed observations from multiple viewing geometries can be used to constrain physical properties like retrograde versus prograde pole orientation and thermal inertia. The FG3 results are more ambiguous with detected thermal variability possibly due to systematic issues with NEATM, an unexpected prograde rotation state, or a surface that is spectrally and thermally heterogenous. This study highlights the potential diagnostic importance of high phase angle thermal measurements on both sides of opposition. We find that the NEATM thermal beaming parameters derived from our near-IR data tend to be of order 10's of percent higher than parameters from ensemble analyses of longer wavelength data sets. However, a systematic comparison of NEATM applied to data in different wavelength regimes is needed to understand whether this offset is simply a reflection of small number statistics or an intrinsic limitation of NEATM when applied to near-IR data. With the small sample presented here, it remains unclear whether NEATM modeling at near-IR wavelengths can robustly determine physical properties like pole orientation and thermal inertia.

\end{abstract}

\begin{keyword}
Asteroids\sep Spectroscopy\sep Infrared observations
\end{keyword}



\section{Introduction}

Obliquity (a scalar) or spin vector is a fundamental property of all celestial bodies. In the Solar System, spin vectors directly influence the efficiency in which the Yarkovsky and YORP effects (i.e. thermal radiation forces and torques) change the orbital and rotational properties of minor planets \citep{rubincam00,bottke06,vok15}. The Yarkovsky effect plays a key role in the dynamical transport of small asteroids from the Main Belt into the near-Earth population \citep{farinella99}. The YORP effect directly influences rotation state and is frequently invoked to explain the formation of binary and multiple asteroid systems through rotational fission or mass shedding \citep{margot02,scheeres07,walsh08,pravec10,jacobson11}. Furthermore, spin states are a direct consequence of asteroid-asteroid impacts and thus trace collisional environments across dynamical populations \citep{paolicchi02}.

Despite the importance of spin vectors, only a few hundred out of three quarters of a million known asteroids have constrained pole orientations. This is a direct consequence of the many years or even decades required to constrain spin vectors through astrometric or photometric observations \citep[e.g.][]{kaasalainen02,slivan02,lowry07,nugent12}. Analyses across shorter temporal baselines often suffer from retrograde-prograde degeneracies. Doppler delay radar imaging can determine shape and pole solutions, but requires sufficient sky motion during the observing window and is limited to the closest near-Earth and largest Main Belt asteroids \citep{ostro02}. 

\begin{figure}[h]
\begin{center}
\includegraphics[width=15cm]{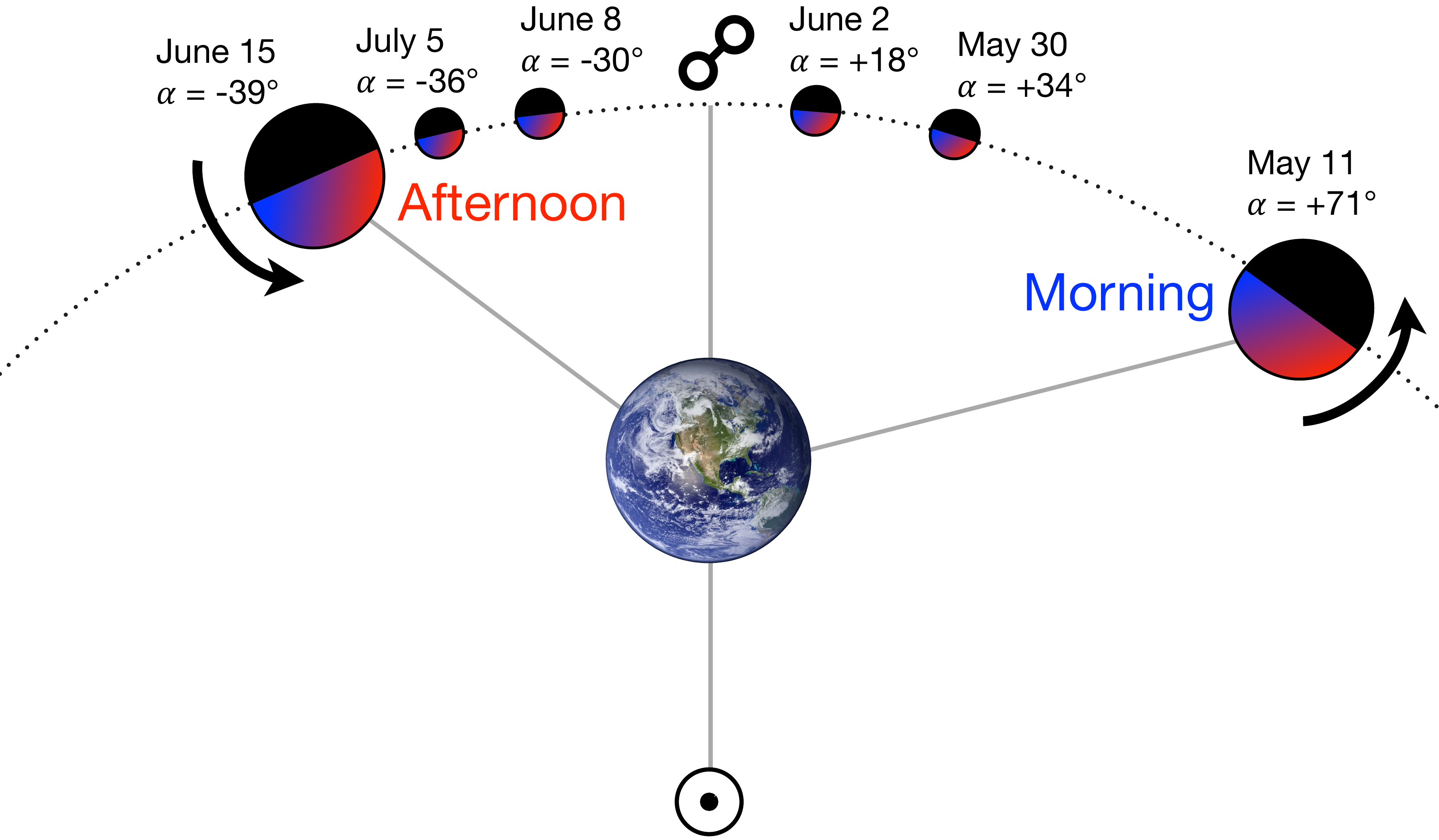}
\end{center}
\caption[]{Schematic of the opposition (\opposition) centered orbital longitude of 1998 QE2 during its 2013 apparition. The dates and corresponding solar phase angles ($\alpha$, signed relative to opposition) are indicated for each of our observations. The asteroid cartoons are shaded to indicate illumination (black = nighttime hemisphere) and surface temperate distribution for a body with finite thermal inertia (blue = cold, red = hot). The predicted prograde rotation of the asteroid \citep{springmann14} is indicated by the black arrows. The cartoons of the asteroid at maximal phase angles ($+71^\circ$ and $-39^\circ$) are larger to emphasize the preferential viewing of cooler morning temperatures prior to opposition and warmer afternoon temperatures post-opposition. Note the phase angle actually decreased after the June 15 observation. Earth image credit: NASA/GSFC.
} 
\label{fig.cartoon}
\label{lastfig}
\end{figure}

Here we focus on an alternate technique for constraining spin orientation, namely by monitoring thermal emission as a function of solar phase angle. This technique does not solve for the obliquity or specific spin vector, but can constrain retrograde versus prograde spin orientation. Measuring thermal emission from a body at multiple viewing geometries on either side of opposition can reveal differences in local morning versus afternoon temperature distributions across the surface of an asteroid (Figure \ref{fig.cartoon}). The rotational sense of the body directly influences the local time and temperature distribution on the asteroid surface as perceived from Earth, particularly when observing at geometries of high phase angle. This afternoon-morning temperature dichotomy can observationally manifest as a difference in thermal emission. For the case depicted in Figure \ref{fig.cartoon}, preferential viewing of the ``cold'' hemisphere before opposition and the ``warm'' hemisphere afterwards would be reversed if this object were in retrograde rotation. Therefore, variations in thermal flux on either side of opposition can serve to constrain pole orientation. Thermal observations across a range of phase angles is a standard method for determining asteroid thermal inertia \citep{delbo15}. 

This technique, originally described \citep[][]{matson71} and pioneered in the 1970's, has not been widely used to constrain pole orientation, but has seen some success. \citet{morrison76} confirmed that the near-Earth asteroid 433 Eros is in a prograde rotation state based on enhanced 10- and 20-$\mu m$ thermal emission prior to opposition, consistent with preferential viewing of local afternoon. Even though Eros' pole only points 11$^\circ$ above the ecliptic \citep{zuber00} its moderate orbital inclination (10.8$^\circ$) and observations by \citet{morrison76} at moderate to high solar phase angles (-41$^\circ$, -24$^\circ$, and +42$^\circ$) ensured that the afternoon and morning sides of Eros were preferentially viewed prior to and after opposition respectively. \citet{morrison77} and \citet{hansen77} extended this technique to several large Main Belt asteroids. They were able to correctly predict prograde rotation for 1 Ceres \citep{johnson83}, 4 Vesta \citep{gehrels67} and 19 Fortuna \citep{torppa03}, and a retrograde state for 10 Hygeia \citep{kaasalainen02}. However they did not correctly predict the retrograde rotation of 2 Pallas \citep{torppa03,carry10}. There are likely several reasons for this. First, the prediction for Pallas was tenuous anyways due to it having the fewest observations of the objects under investigation. Second, the standard thermal model used to derive physical properties (including pole orientation) assumed non-rotating spherical bodies. Third, the range of accessible phase angles for objects in the Main Belt is limited relative to near-Earth objects. Finally, the possibility of low surface thermal inertia for large ($>100$ km) asteroids \citep[e.g.][]{delbo15} acts to dampen morning-afternoon temperature gradients. In light of these issues, it is impressive that these early investigations resulted in meaningful constraints on the rotation states of any of the targeted asteroids. Since these initial studies, the use of a morning-afternoon temperature dichotomy to inform pole orientation has only been discussed and applied in a few cases  \citep[e.g.][]{lebofsky86,harris07,muller12}. Note that this method is not  sensitive to objects with poles aligned near to the line of sight, nor does it provide a measure of the object obliquity, instead it simply provides a means to constrain a prograde versus retrograde spin state.

Traditionally, the measurement of asteroid thermal emission and subsequent modeling of physical properties has relied on mid-infrared (3-25 $\mu m$) photometry. An asteroid thermo-physical model called SHERMAN has been developed for application to airless bodies of arbitrary shape and has primarily been applied to (though is not restricted to) observations at wavelengths between 2.5 - 4.0 $\mu$m \citep{howell15}. However, low albedo ($<10\%$) asteroids at small heliocentric distances ($<1.2$ AU) have high enough surface temperatures that the Wien tail of their thermal emission profiles can be detected at near-infrared wavelengths (0.8 - 2.5 $\mu$m). At such short wavelengths the measured signal is a combination of solar reflectance and thermal emission. This thermal excess or thermal tail can be modeled to retrieve surface temperature and albedo \citep{rivkin05,reddy09,reddy12}. However, models of this excess from normalized reflectance spectra are not directly sensitive to diameter because such data are not flux calibrated.

Here we present case studies of the near-Earth asteroids (285263) 1998 QE2 ($\S3$) and (175706) 1996 FG3 ($\S4$) based on observations during apparitions favorable to the detection of phase-dependent thermal variability. This analysis represents the first systematic study of thermal variability in the near-IR K-band as a function of phase angle.

\section{Observations and Thermal Modeling}

Both 1998 QE2 and 1996 FG3 have low geometric albedos of about 3\% and 4\% respectively \citep{trilling10,wolters11}, and thus display significant thermal emission in the near-infrared when they are at heliocentric distances of $\sim1$ AU. We present near-infrared spectra of 1998 QE2 taken between May and July of 2013 and of 1996 FG3 taken during apparitions in 2009 and 2011 (Tables \ref{tab.QE2obs} and  \ref{tab.FG3obs}). All data analyzed here were obtained with SpeX at NASA's Infrared Telescope Facility, which was configured in its low resolution (R $\sim$ 100) prism mode with a 0.8" slit \citep{rayner03}. All observations were conducted with the telescope operating in a standard ABBA nod pattern. Commonly-used solar analogs \citep{demeo09} were observed to correct for telluric absorption and to remove the solar contribution from the measured reflectance. Reduction protocols followed \citet{moskovitz13} and \citet{demeo09}. 

\begin{center}
\begin{table}[]
\begin{tabular}{lccccccc}
\hline 
\hline
Observation Date	& $V$		& $r$	& $\Delta$	& $\alpha$		& $\eta_{3\%}$ 			& $\eta_{6\%}$			& $\eta_{1\%}$\\
(UT)				& [mag]		& [AU]	& [AU]		& [deg]		&  					&					& \\
\hline
2013 May 11 		& 15.4		& 1.045	& 0.129		& +71.0		& 1.56 $\pm$ 0.05 		& 1.29				& 2.03  \\
2013 May 30 		& 11.8		& 1.047	& 0.040		& +34.0		& 1.18 $\pm$ 0.02		& 0.97 				& 1.57 \\
2013 June 2 		& 11.4		& 1.052	& 0.040		& +17.9		& 1.08 $\pm$ 0.03		& 0.89 				& 1.45  \\
2013 June 8 		& 12.7		& 1.067	& 0.061		& -30.4		& 1.11 $\pm$ 0.02		& 0.91				& 1.48  \\
2013 June 15 		& 14.0		& 1.091	& 0.099		& -38.8		& 1.16 $\pm$ 0.03		& 0.95 				& 1.54  \\
2013 July 5 		& 15.9		& 1.189	& 0.223		& -35.9		& 1.10 $\pm$ 0.03		& 0.91 				& 1.45  \\
\hline
\hline
\end{tabular}
\caption[]{Summary of (285263) 1998 QE2 observations and best-fit NEATM thermal beaming parameter ($\eta$) values for the full range of possible albedos from 1-6\%. 
}
\label{tab.QE2obs}
\end{table}%
\end{center}

\begin{landscape}
\begin{center}
\begin{table}[h]
\begin{tabular}{lcccccccl}
\hline 
\hline
Observation Date	& $V$		& $r$	& $\Delta$	& $\alpha$		& $\eta_{3.9\%}$ 		& $\eta_{5.1\%}$		& $\eta_{2.7\%}$	&  \\
(UT)				& [mag]		& [AU]	& [AU]		& [deg]		&  					&					& 				& Mutual Event?\\
\hline
2009 March 30		& 16.2		& 1.223	& 0.227		& +8.3		& 1.35 $\pm$ 0.05 		& 1.27 				& 1.48  			& No \\
2009 April 27		& 16.7		& 1.081	& 0.159		& +58.4		& 1.9 $\pm$ 0.1 		& 1.78 				& 2.07  			& Secondary occulted by primary \\
2011 December 1	& 15.6		& 1.050	& 0.108		& -51.5		& 1.35 $\pm$ 0.05 		& 1.28 				& 1.50  			& Secondary occulted by primary \\
2011 December 30	& 16.4		& 1.203	& 0.228		& +14.0		& 1.12 $\pm$ 0.03		& 1.05 				& 1.23  			& Primary eclipsed and occulted by secondary \\
\hline
\hline
\end{tabular}
\caption[]{Summary of (175706) 1996 FG3 observations and best-fit NEATM $\eta$ values for the full range of possible albedos from 2.7-5.1\%.
}
\label{lasttable}
\label{tab.FG3obs}
\end{table}%
\end{center}
\end{landscape}

We model the thermal tails of our targets with the near-Earth asteroid thermal model \citep[NEATM;][]{harris98}. A number of parameters go into this model including slope parameter ($G$), visible to near-IR reflectance ratio (0.55 $\mu$m / 2.5 $\mu$m), emissivity, and the coupled parameters geometric albedo ($p_V$), diameter, and absolute magnitude ($H$). The NEATM assumes that the asteroid is spherical and that the hottest point on the surface is the sub-solar point (i.e. the surface is in instantaneous equilibrium with the incident solar flux). The NEATM thermal beaming parameter $\eta$ compensates for a lack of explicit treatment of thermophysical properties like thermal inertia and surface roughness. In the NEATM, $\eta$ is inversely proportional to temperature to the fourth power\footnote{Technically, in NEATM formalism $\eta T_{SS}^4$ is a constant, where $T_{SS}$ is the sub-solar equilibrium temperature on the surface of the asteroid.} and thus can indicate whether the measured thermal emission is higher or lower than expected from an idealized, smooth surface with zero thermal inertia at zero phase angle. Previous application of the NEATM to multi-epoch observations has successfully produced results that are broadly consistent with those from more sophisticated thermophysical models \citep[e.g.][]{mueller06,harris07,delbo15}.

\citet{delbo04} presented a comprehensive suite of NEATM models that quantitatively describe thermal variability as a function of solar phase angle as well as other physical parameters. These models suggest several possible caveats to the detectability of aspect-dependent thermal variability. 
\begin{enumerate}
\item The dimensionless thermal parameter $\Theta$ \citep{spencer89} quantifies the ratio of radiative to rotational timescales, and is a function of spin period, thermal inertia ($\Gamma$), and subsolar temperature as dictated by heliocentric distance. Though the smallest near-Earth asteroids can spin rapidly, the vast majority of NEAs with sizes bigger than about 200 meters have rotation periods on the order of several hours \citep[e.g.][]{polishook12}. Thus, for typical NEA rotation periods and heliocentric distances ($\sim1$ AU), a thermal inertia greater than about 200 in MKS units (J m$^{-2}$ s$^{-0.5}$ K$^{-1}$) ensures that heat deposited by solar insolation does not radiate away so quickly as to prevent morning-afternoon temperature differences. The typical thermal inertia for asteroids with diameters $<5$km would be in that range \citep{delbo07}. While high thermal inertia certainly facilitates a morning-afternoon temperature dichotomy, thermally-constrained poles for several large Main Belt asteroids \citep{morrison77,hansen77} suggests that either these early determinations were simply lucky or that high thermal inertia is not an exclusive requirement.

\item The asteroid must be observable across a wide range of solar phase angles, particularly at phases in excess of $\sim50^\circ$. In general, such phase angle ranges are only accessible for NEAs. 

\item If thermal emission is to be detected in the near-IR K-band, then the object needs to be at a heliocentric distance inside of about 1 AU and have an albedo $<10\%$. 

\item The asteroid should be observed during a single apparition to avoid significant changes in viewing aspect that could occur across multi-epoch observations.
\end{enumerate}

Even though all of these properties were not necessarily known in advance for QE2 and FG3, their low albedos and wide range of observable phase angles made them good candidates for monitoring thermal variability in the near-IR as a function of viewing aspect.

\section{Case Study: (285263) 1998 QE2}

\begin{figure}[]
\begin{center}
\includegraphics[width=12cm]{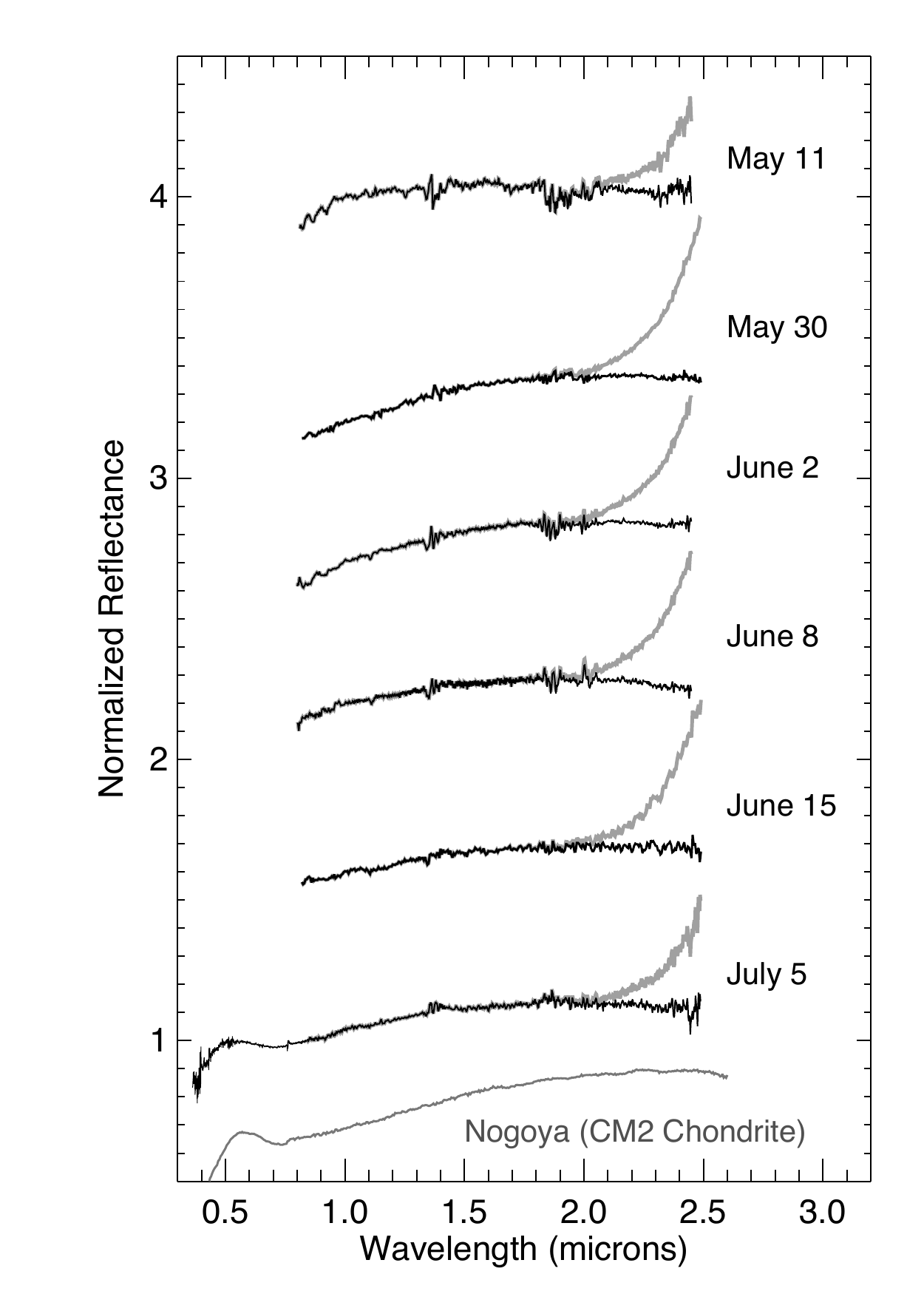}
\end{center}
\caption[]{IRTF/SpeX spectra of 1998 QE2 taken in the months surrounding its opposition passage in mid 2013. The asteroid spectrum from July 5 also includes visible data (0.4 - 0.8 $\mu$m) from \citet{hicks13}. QE2 is taxonomically classified as a Ch-type \citep{demeo09} and is a close spectral analog to the CM2 carbonaceous chondrite Nagoya. The measured spectra are shown in grey, the NEATM-corrected spectra with thermal tails removed are overplotted in black. Spectra have been vertically offset for clarity.
} 
\label{fig.spec}
\label{lastfig}
\end{figure}

Radar observations of 1998 QE2 in 2013 revealed a diameter of approximately 3 km and a prograde rotation state with the pole pointing approximately 40$^\circ$ above the ecliptic\footnote{Incidentally these observations also discovered a small satellite roughly 750 meters in size. Unless the satellite has a dramatically different albedo or thermal properties relative to the primary, its contribution to the system's measured flux should not significantly influence any of the results presented here.} \citep{springmann14}. From the JPL Horizons system the absolute magnitude of QE2 is $H=17.3$, which is revised from a value of $H=16.4$ used in previous works \citep[e.g.][]{trilling10}. An updated analysis of warm Spitzer observations using this new H magnitude yields a low albedo for QE2 of $3_{-2}^{+3}\%$ \citep[][]{trilling16}. We adopt $H=17.3$, even though it may be less accurate than the value of $H=16.98$ measured by \citet{hicks13}, because it is fully consistent with the formal log-normal albedo error bars as computed by \citet{trilling16}. As we discuss below, these are important for investigating systematic errors in our NEATM models due to uncertainties on the input parrameters. Figure \ref{fig.cartoon} depicts a schematic of the opposition-centered orbital geometry of QE2 during its 2013 apparition. To facilitate our analysis we employ a signed phase angle ($\alpha$) based on the geocentric ecliptic longitude of asteroid opposition, specifically $\alpha < 0^\circ$ when the asteroid's ecliptic longitude minus the Sun's ecliptic longitude is greater than 180$^\circ$. The putative prograde rotation of QE2 suggests that prior to opposition ($\alpha > 0^\circ$) Earth-based observers would preferentially see the cooler morning side of the asteroid, whereas post-opposition ($\alpha < 0^\circ$) observations would preferentially access the warmer afternoon side. Near-IR spectra of QE2 spanning opposition are shown in Figure \ref{fig.spec}.

A composite visible/near-infrared spectrum of QE2 is shown at the bottom of Figure \ref{fig.spec}. Due to the presence of a weak 0.7 $\mu$m absorption feature the asteroid is best fit by a Ch-type taxonomy in the \citet{demeo09} system. Using the technique outlined in \citet{moskovitz13} we find that QE2 has a spectrum and albedo most closely matched to RELAB \citep{pieters83} spectra of carbonaceous chondrites. The CM2 Nagoya is a particularly good spectral analog (Fig. \ref{fig.spec}). A potential link between Ch-asteroids and CM chondrites has been noted previously \citep[e.g.][]{burbine98}. This taxonomic classification is consistent with the low Spitzer albedo.

The spectra in Figure \ref{fig.spec} clearly show thermal emission long-wards of $\sim2 \mu$m. We model these thermal tails with the following NEATM parameters: geometric albedo $p_V$ = $3_{-2}^{+3}\%$ \citep[][]{trilling16}; slope parameter G = 0.086, equal to the mean for C-complex asteroids \citep[][]{warner09}; visible to near-IR reflectance ratio = 1.11, measured from our combined visible and near-IR spectral data (Fig. \ref{fig.spec}); absolute magnitude H = 17.3 from the JPL Horizons system; and emissivity $\epsilon=0.9$. These values are either consistent with observations of QE2 or are typical for objects like QE2, and are held constant for all of the modeling conducted here. Again we adopt $H$ from JPL as opposed to \citet{hicks13} so that we can propogate the albedo error bars from \citet{trilling16} who also used the JPL value.

With each of these parameters held fixed, the beaming parameter $\eta$ is left as a free parameter. We vary $\eta$ to best remove the thermal tails from the spectra in Figure \ref{fig.spec}. The best fit $\eta$ for each spectrum is determined by minimizing the error-weighted RMS residual between the thermally corrected spectrum and a linear fit to the data between 1.9 and 2.49 $\mu$m. This approach thus requires three free parameters: $\eta$ and the two parameters associated with the linear fit. Uncertainties on the fitted $\eta$ values reflecting the signal-to-noise of the data (i.e. statistical errors) are conservatively determined by bracketing the range of $\eta$ values that clearly over- or under-compensate the thermal emission. These uncertainties are somewhat subjective but do correspond to deviations from the best fit by approximately one-sigma as dictated by the noise level in the spectral data. Best-fit $\eta$ values and signal-to-noise related uncertainties are given in Table \ref{tab.QE2obs} for the nominal assumed albedo of 3\%. A similar approach to modeling of near-IR spectra and estimating uncertainties on the model parameters has been employed elsewhere \citep[e.g.][]{rivkin05,reddy12}.

We further investigate uncertainty on $\eta$ by considering systematic errors on the NEATM input parameters which would produce an overall offset to the best-fit $\eta$ values. The assumed albedo is the dominant source of systematic uncertainty in $\eta$, which for QE2 ranges from 1-6\% \citep{trilling16}. We fit $\eta$ assuming these extreme values and include those additional fits in Table \ref{tab.QE2obs}. The uncertain albedo for QE2 results in large systematic errors ($\sim50\%$) in the absolute $\eta$ values. However, any change in the albedo from the nominal 3\% affects all $\eta$ values in the same way (increase or decrease) and by approximately the same amount. The assumed value of G = 0.086 also represents a potential source of systematic error. However, considering a wide range of G values from -0.1 to +0.15 results in $\eta$ fits that span less than half the range produced by considering the full extent of possible albedos. In other words the systematic errors induced by uncertain G are approximately 20\% on the best fit $\eta$ values. 

\begin{figure}[t]
\begin{center}
\includegraphics[width=15cm]{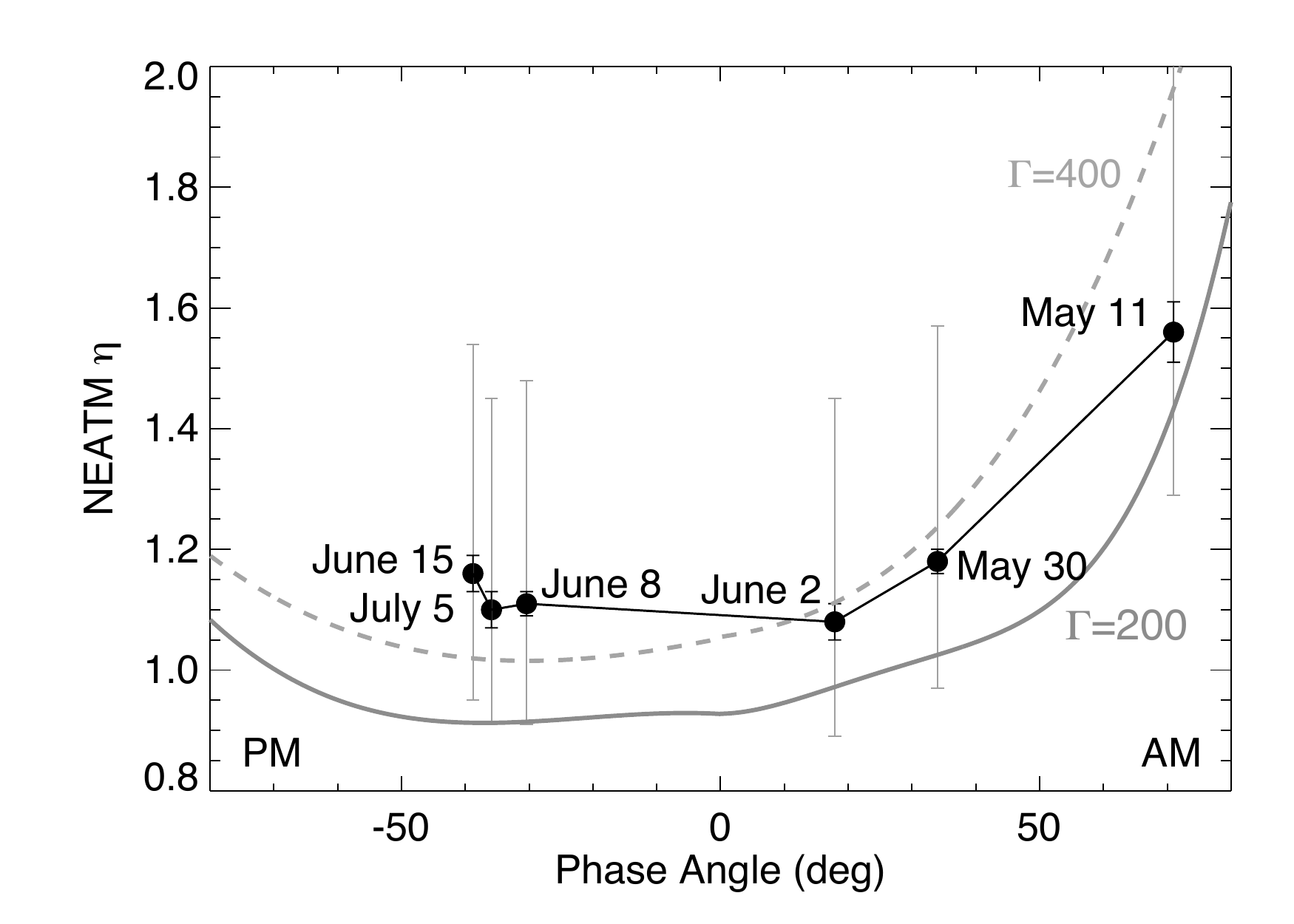}
\end{center}
\caption[]{NEATM $\eta$ as a function of solar phase angle, $\alpha$, for 1998 QE2 during its 2013 apparition. As QE2 moved through opposition it tracks from the upper right corner to the lower left of this $\eta-\alpha$ space (black circles). NEATM models for asteroids with prograde rotation, surface roughness $\bar{\theta}=36^\circ$, and thermal inertia ($\Gamma$) of 200 and 400 in MKS units are over-plotted as the solid and dashed grey lines respectively \citep{delbo04}. For a prograde rotator the predicted $\eta-\alpha$ relationship is asymmetric with higher $\eta$ (lower morning temperatures) at positive phase angles and lower $\eta$ (higher afternoon temperatures) at negative phase angles. The curves for retrograde rotators would be flipped about $\alpha=0^\circ$. Statistical errors based on the signal-to-noise of our data are in black, while systematic error bars accounting for uncertainty in albedo are shown in grey.  
} 
\label{fig.eta}
\label{lastfig}
\end{figure}

The NEATM results for QE2 are shown in Figure \ref{fig.eta}. An asymmetric trajectory in this $\eta-\alpha$ parameter space can indicate physical properties such as sense of rotation or thermal inertia, however the use of NEATM to diagnose these physical traits must be treated with caution due to known systematic errors inherent to NEATM \citep{wolters09} and known correlations between $\eta$ and $\alpha$ \citep{delbo03,wolters08,masiero11}. Nevertheless, the results in Figure \ref{fig.eta} are suggestive that a small number of carefully timed observations in the near-infrared, particularly at phase angles $>50^\circ$, could be used to constrain physical properties of small NEAs. Unfortunately the lack of repeat observations of QE2 at high phase angles means that the results for this specific object are tenuously hinged to the single May 11 data point.

In consideration of this technique, we compare our QE2 results to model curves extracted from \citet{delbo04}. Specifically, we compare our $\eta$ values to a suite of models that correspond to bodies with a macroscopic surface roughness parameter $\bar{\theta}=36^\circ$. This particular value was selected as an intermediate case in the range of models presented by \citet{delbo04} and is consistent with expected roughness parameters from more sophisticated photometric models of other well studied asteroids \citep[e.g.][]{delbo09}. The assumption of this particular roughness does not significantly influence the results presented here. If we were to adopt other roughness values (10$^\circ$, 20$^\circ$, 58$^\circ$) then the curves in Figure \ref{fig.eta} would shift by no more than $\sim10-20\%$. More specifically, the afternoon-side curves ($\alpha<0^\circ$) remain largely unchanged across all roughness values, whereas the morning-side ($\alpha>0^\circ$) curves either get steeper by $\sim20\%$ for the largest roughness or shallower by $\sim10\%$ for the smallest roughness value.

As seen in Figure \ref{fig.eta}, the large $\eta$ prior to opposition (May 11) suggests a cooler relative surface temperature, consistent with (though not uniquely diagnostic of) viewing of local morning on the asteroid surface and hence prograde rotation. Furthermore, the $\eta$ values and magnitude of the $\eta$-enhancement at high positive phase angles are consistent with models for a prograde rotator with thermal inertial $\Gamma\sim200-400$ \citep{delbo04}. Interestingly, a thermal inertia around 200 is the expectation for a $\sim3$-km asteroid like QE2 \citep{delbo15}. Again, the assumption of surface roughness $=36^\circ$ does not significantly affect the rough fit to our data by theoretical curves corresponding to thermal inertia values around 200-400. For the full range of roughness values considered by \citet{delbo04}, a thermal inertia of 50 is too low to produce large variations in $\eta$, and thermal inertia of 900 results in values of $\eta$ that are much larger than those retrieved from our data. The systematic uncertainty in our $\eta$ values due to the range of possible albedos could shift the points in Figure \ref{fig.eta} up or down by several tens of percent, however the general result that the data are indicative of a thermal inertia of a few hundred would still hold. Furthermore, the relative orientation and increase at high positive phase angle of the $\eta-\alpha$ points in Figure \ref{fig.eta} would be preserved.

\section{Case Study: (175706) 1996 FG3}

The NEA 1996 FG3 is also a binary system \citep{pravec00} with a primary about 1.7 km in size \citep{wolters11,mueller11,walsh12,yu14}, a spacecraft accessible orbit (i.e. $\Delta v=6.6$ km/s), and a putative primitive composition \citep{binzel01,deleon11,walsh12,deleon13}. FG3's satellite is roughly 450 meters in size \citep{yu14} and is in a tidaly-locked retrograde orbit \citep{benner12,scheirich15}, which suggests that the primary is also in a retrograde spin state. Again, the satellite is unlikely to have a significant influence on the system's thermal flux unless it has a dramatically different albedo or thermal properties relative to the primary. A low albedo around 4\% \citep{wolters11} makes FG3 another ideal object for monitoring thermal variability in the near-IR. As with QE2, the flux contribution from the small satellite should not have a significant effect on any of the measurements presented here.

\begin{figure}[]
\begin{center}
\includegraphics[width=15cm]{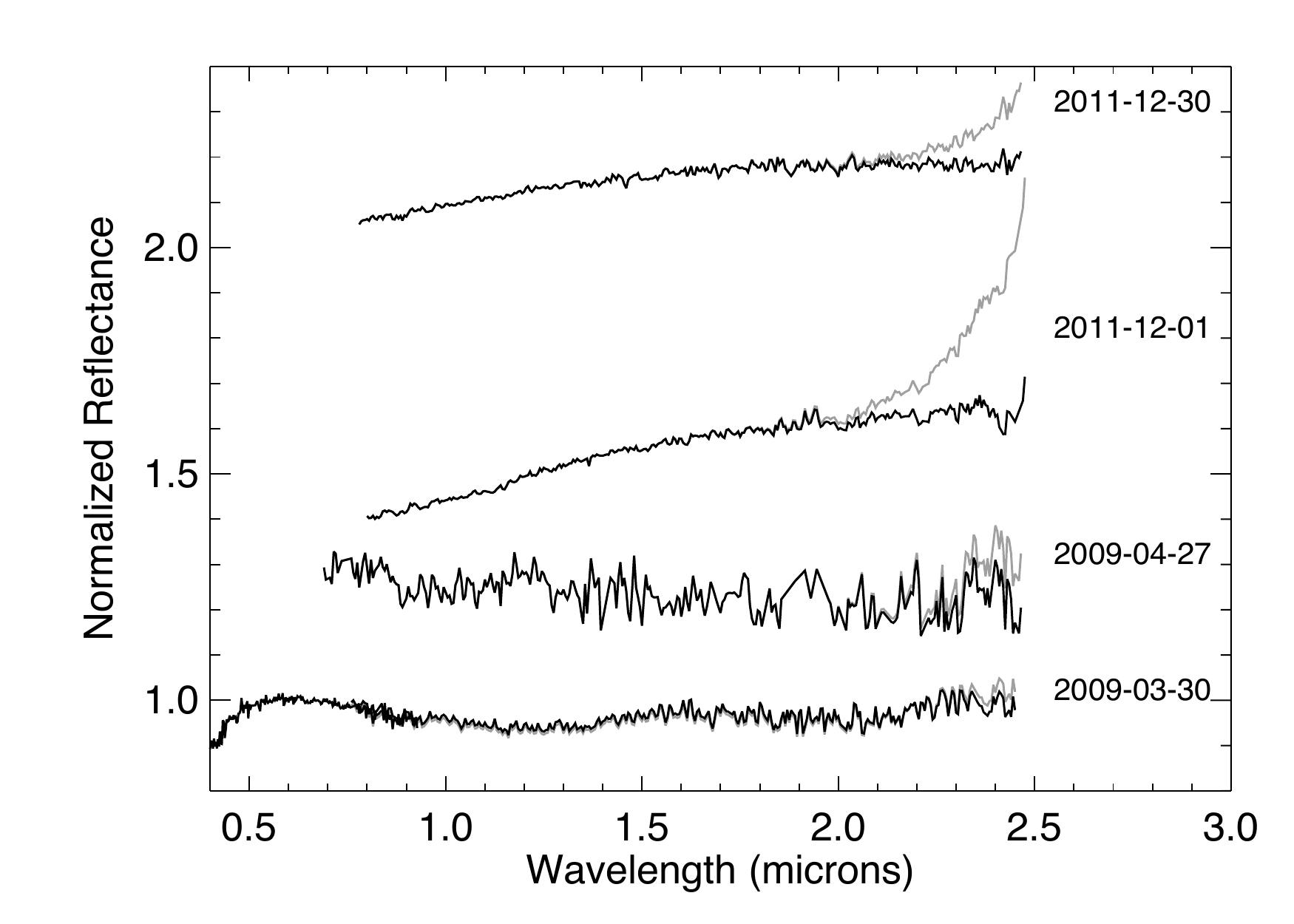}
\end{center}
\caption[]{Reflectance spectra of 1996 FG3 (grey lines), offset for clarity. In general the data from 2009 display a neutral spectral slope and limited thermal emission, whereas the data from 2011 show a red spectral slope and pronounced thermal emission. Visible data from \citet{binzel01} have been appended to the bottom spectrum. NEATM corrected spectra are over-plotted in black.
} 
\label{fig.fg3spec}
\label{lastfig}
\end{figure}

Numerous near-IR spectra of FG3 have been obtained during apparitions in 2009 and 2011 \citep{deleon11,walsh12,deleon13}. We analyze four individual spectra from these apparitions (Table \ref{tab.FG3obs}; Figure \ref{fig.fg3spec}). These spectra can be qualitatively divided into two groups: in 2009 the spectra have a neutral spectral slope and minimal thermal excess, whereas in 2011 the spectra show a pronounced red spectral slope and clear thermal emission. Additional near-IR spectra not analyzed here \citep[][]{deleon11,rivkin12} are consistent with this qualitative description. Taxonomically \citep{demeo09}, the 2009 data suggest that FG3 is a member of the C-complex while the 2011 data suggest membership in the X-complex. The repeatability of these spectra on different nights in each of the 2009 and 2011 epochs suggests that this variability is real and not an observational artifact. 

We have compared the timing of these observations to predictions for mutual events (P. Scheirich, private communication). Those predictions are included in Table \ref{tab.FG3obs}. In short, the occurrence of mutual events does not seem to have any major influence on the spectral data, which is not surprising considering the significant size difference between the primary and secondary. The two spectra that differ the most, i.e. the 2009 April 27 and 2011 December 1 data, were both obtained when the secondary was occulted by the primary. Furthermore, within each epoch the pairs of spectra were obtained during different system configurations, again suggesting that the secondary had little optical and thermal influence on the measurements. It is worth noting however that occultation of the primary by the secondary, particular if the secondary shadows the hottest region on the primary's surface, could have significant thermal implications. This occultation would lead to reduced thermal emission and a correspondingly elevated $\eta$. The only data analyzed here that corresponded to occultation of the primary were from 2011 December 30. This data point is further discussed below.

\begin{figure}[h]
\begin{center}
\includegraphics[width=15cm]{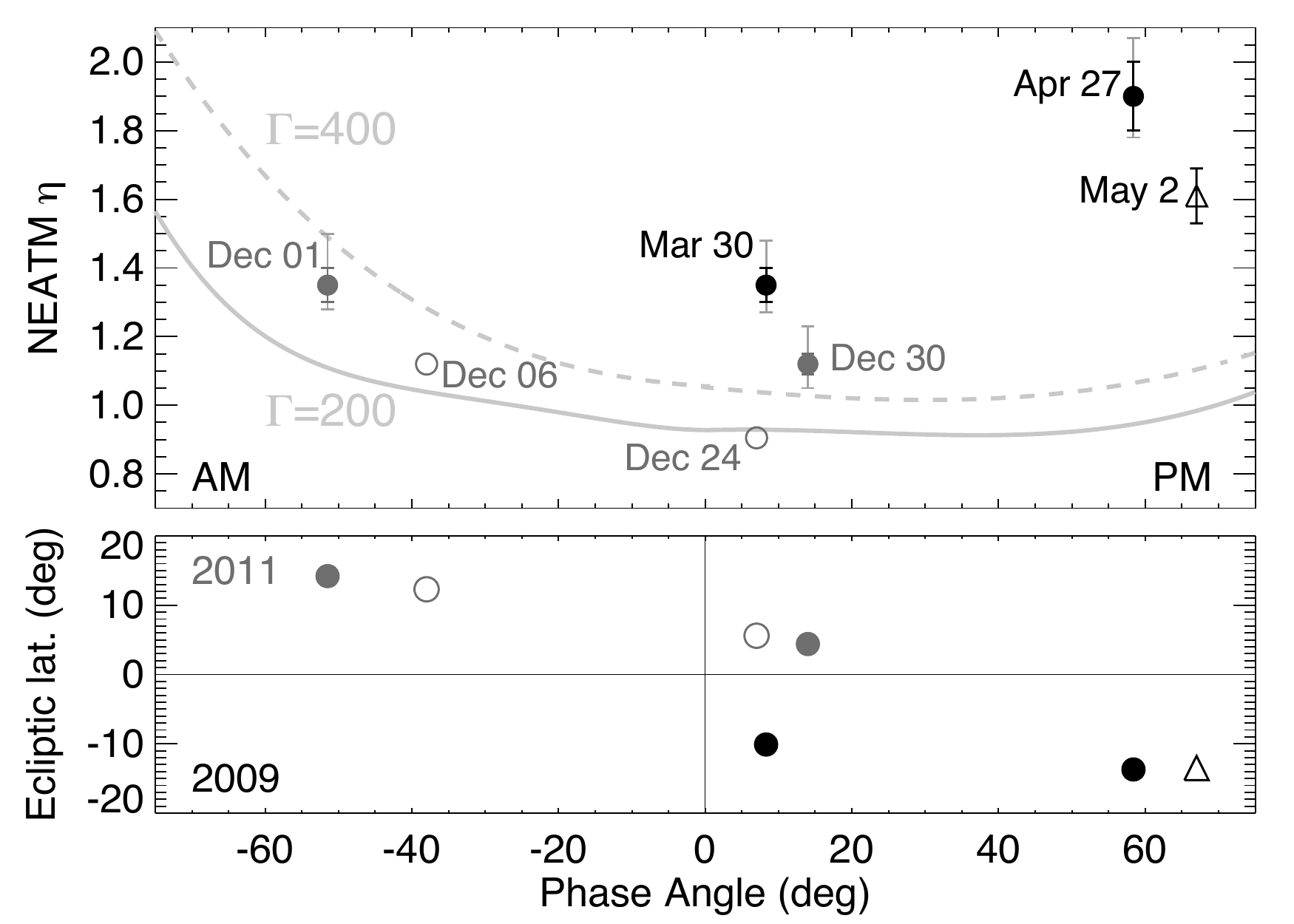}
\end{center}
\caption[]{Top panel: The $\eta-\alpha$ relationship for 1996 FG3 based on 2009 (black) and 2011 (grey) observations. Filled circles are the data analyzed here (Figure \ref{fig.fg3spec}). Statistical errors based on the signal-to-noise of the data are shown in black, systematic errors due to the range of possible albedos are in grey. Open circles are from \citet{rivkin13} and the open triangle is from \citet{walsh12}. NEATM models for a retrograde rotator with thermal inertia ($\Gamma$) of 200 and 400 in MKS units, and surface roughness $\bar{\theta}=36^\circ$ are over-plotted \citep{delbo04}. The curves for prograde rotators would be flipped about $\alpha=0^\circ$. The 2011 data are broadly consistent with the models for retrograde rotation, however the 2009 data are not. Bottom panel: the geocentric ecliptic latitude of the asteroid at the time of each observation. The discrepancy between the data and the models could be explained by a texturally and spectrally heterogeneous surface revealed by the different viewing geometries, i.e. preferential viewing of the southern hemisphere in 2011 and of the northern hemisphere in 2009.
} 
\label{fig.fg3therm}
\label{lastfig}
\end{figure}

As with 1998 QE2 (\S3) we model the thermal tails with the NEATM. The following parameters are employed: geometric albedo $p_V$ = $3.9 \pm 1.2\%$ \citep[][]{walsh12}; absolute magnitude H = $17.76 \pm 0.03$ and slope parameter G = $-0.07\pm 0.02$ as measured by \citet[][]{pravec06}; visible to near-IR reflectance ratio ranging from 0.9-1.2 as measured from our combined visible and near-IR spectra; and emissivity $\epsilon=0.9$. These values are held constant for all of the FG3 models and again $\eta$ is left as a free parameter in the NEATM fits. Our NEATM results for FG3 are shown in Figure \ref{fig.fg3therm}. As with QE2 we compute error bars on the derived $\eta$ values that account for uncertainty due to the signal-to-noise of the data and we compute systematic offsets in $\eta$ attributable to the range of possible albedos for FG3 \citep{walsh12}. Again the uncertainty on albedo is the dominant source of error for these models. Previous results based on observations at longer wavelengths are also included in Figure \ref{fig.fg3therm}; in particular, NEATM $\eta$ values from $2-4~\mu m$ spectral observations by \citet{rivkin13} and thermal infrared ($\sim8-20~\mu m$) photometry by \citet{walsh12}.

Unlike the $\eta-\alpha$ results for 1998 QE2, the fits in Figure \ref{fig.fg3therm} do not clearly follow the NEATM predictions for a retrograde rotator. In fact the highest values of $\eta$ for FG3 correspond to viewing geometries that would have preferentially viewed the afternoon hemisphere for a retrograde rotator. Our data thus suggest that FG3 is in a prograde rotation state. This apparent contradiction is discussed in the following section. Values of $\Gamma \sim 200-400$ most closely overlap our data points and are broadly consistent, though higher by a factor of $\sim2$, with other estimates for FG3's surface \citep{wolters11}. Interestingly the  $\eta$ value for December 30 is slightly higher than expected for a low phase angle observation. As mentioned above this could be attributed to observation of the system while the primary was occulted by the secondary. Modeling of this scenario is beyond the scope of this work, and given the unexplained variability in $\eta$ for this object we do not pursue this idea any further.

\section{Discussion \label{sec.disc}}

We have demonstrated the detection of phase dependent variations in thermal beaming parameter for the NEAs (285263) 1998 QE2 and (175706) 1996 FG3. While such variability is not novel -- correlations between phase angle and $\eta$ have been shown previously \citep[e.g][]{delbo03,wolters08,masiero11} -- this work represents the first systematic analysis of such variability at near-IR wavelengths. Ultimately the goal of any such investigation would be to correlate actual physical properties with derived model parameters. It is in this vein that we discuss the thermal variability seen for QE2 and FG3.

As discussed above, the NEATM modeling of the QE2 data indicate a significant increase in $\eta$ at high positive phase angles (Figure \ref{fig.eta}). This is reasonably well fit with models for a prograde pole orientation and a surface thermal inertia of roughly 200-400 J m$^{-2}$ s$^{-0.5}$ K$^{-1}$, properties which are consistent with other observations \citep{springmann14} and expectations for objects of this size \citep{delbo07}. It is unclear whether the pole, pointing at ecliptic latitude $\sim+40^\circ$ is near enough to purely prograde to fully explain the observed thermal effects as a function of solar phase angle.

While our QE2 results are suggestive that the thermal variability in the near-IR is indicative of physical properties, a data point at large negative phase angles ($\alpha<-50^\circ$) would have helped to definitively determine whether the apparent asymmetry in this $\eta-\alpha$ space is a signature of pole orientation or simply a side effect of the NEATM method. The diagnostic importance of such high phase angle measurements on both sides of opposition is an important lesson for guiding future observational campaigns. In general, the models of \citet{delbo04} suggest that objects with $\Gamma<<200$ would not display significant differences in $\eta$ as a function of phase angle, however future targeting of low albedo NEAs at near-IR wavelengths across a range of sizes are needed to test this model prediction.

The interpretation of the FG3 data is less clear, which is perhaps unsurprising given the numerous previous attempts to understand its variability in reflectance and thermal emission \citep{deleon11,binzel12,deleon13}. We have built upon these previous works and show that there are clear differences in $\eta$ as a function of phase angle (Figure \ref{fig.fg3therm}), and that those differences are asymmetric about opposition. We argue that these differences are not an artifact of any observational issues such as signal-to-noise or data calibration. The 2009 data are of lower signal-to-noise relative to the 2011 data and they do show the greatest offsets from model predictions (Fig. \ref{fig.fg3therm}), but there are no clear indicators of poor calibration. For example, the near complete removal of any residual telluric absorption features around 2.0 $\mu$m suggests that the dispersion solution is good and that the atmosphere was stable enough during those nights to enable removal of the telluric bands. Furthermore, there are no odd continuum or slope changes in the 2009 data that might indicate use of a bad solar analog. Certainly, the repeatability of spectral properties in each of the two epochs suggests that the spectral changes are real. Furthermore, the results of \citet{walsh12} and \citet{rivkin13} are in broad agreement with our NEATM $\eta$ values (Fig. \ref{fig.fg3therm}), thus providing further support for a scenario in which spectral and thermal changes did occur. However, these detected changes, particularly the $\eta-\alpha$ relationship, are not consistent with an expected retrograde spin state \citep{scheirich15,yu14}. As such, we offer three possible explanations for why our results may not be consistent with retrograde models, however in all three cases additional data are needed to better understand the unusual spectral and thermal behavior displayed by FG3.

First, the variability in the data could be a consequence of the known correlation between $\eta$ and phase angle in NEATM models \citep{spencer89,wolters09,reddy12}. Values of $\eta$ derived from NEATM reflect the observed surface temperature. At high phase angles surface roughness causes increased shadowing and an apparent drop in surface temperature (i.e.~higher $\eta$). Such correlations are typically represented as linear fits to an ensemble of data. Table \ref{tab.eta} presents a compilation of coefficients based on linear fits to various NEA data sets. Also included are fit parameters to the FG3 data from the 2009 and 2011 apparitions with errors indicating $1\sigma$ uncertainties. These linear fits are error weighted by the systematic uncertainties on each of the $\eta$ values. In comparison to the ensemble fits our FG3 values are all shifted to higher $\eta$, i.e. larger intercepts, and shallower slopes. The $\eta$ values from the QE2 data are also slightly higher than the mean values from the studies represented in Table \ref{tab.eta}. Though it is not entirely correct to compare our two objects to the diverse ensemble of properties (surface roughness, thermal inertia, morphology, size, spin vector) represented by the population studies in Table \ref{tab.eta}, it is interesting to consider that these offsets may be due to currently unquantified differences in the $\eta$ values returned from fitting near-IR data ($<2.5\mu$m) versus those obtained by fitting longer wavelength data ($>2.5\mu$m). If this is the case then our results provide insight into the magnitude of these offsets -- of order several 10's of percent. However, our limited sample makes it difficult to determine whether this offset is simply a consequence of small number statistics or due to phenomena not treated in the model such as wavelength dependent emissivity.

\begin{table}[h]
\caption{NEATM $\eta-\alpha$ linear-fit coefficients}
\begin{tabular}{lll}
\hline
Reference			&	Intercept			& Slope \\
\hline
\citet{delbo04}		&	$0.92 \pm 0.07$	& $0.011 \pm 0.002$ \\
\citet{wolters08}	&	$0.91 \pm 0.17$	& $0.013 \pm 0.004$ \\
\citet{mainzer11}	&	$0.761 \pm 0.009$	& $0.00963 \pm 0.00015$ \\
This work; FG3 from 2009		&	$1.33 \pm 0.12$	& $0.005 \pm 0.002$ \\
This work; FG3 from 2011		&	$0.94 \pm 0.01$	& $-0.0047 \pm 0.0005$ \\
\hline
\end{tabular}
\label{tab.eta}
\end{table}%

A second explanation for the mismatch between our data and the retrograde models is the unlikely possibility that FG3 is actually in a prograde state. The analysis of light curve photometry strongly suggests that FG3's satellite is in a retrograde orbit \citep{scheirich15} and the satellite seems to be tidally locked \citep{benner12}. Attempts to model the spin pole of the primary have most recently been inconclusive \citep{scheirich15}, but previous analyses suggest a retrograde state \citep{yu14}. While a retrograde satellite orbit and a prograde primary spin state seem unlikely from the perspective of binary formation models involving YORP spin-up and mass loss \citep[e.g.][]{jacobson11}, the current ambiguity in the interpretation of the lightcurve data leaves this as an open possibility. Future observations to unambiguously determine the primary spin state would directly test this scenario.

Interestingly, the linear fits to our FG3 data are significantly different for each of the two epochs of observation (Table \ref{tab.eta}). Furthermore, the reflectance spectra of FG3 are distinctly different in each of the two epochs (Figure \ref{fig.fg3spec}). Perhaps the most interesting comparison to highlight these differences is between the 2009 April 27 and 2011 December 1 data. These data clearly fall into the categories of neutral slope, low thermal emission in 2009, and red slope, high thermal emission in 2011. However the heliocentric distance of FG3 was almost identical on these two dates. The primary difference between these observations is viewing geometry: the solar phase angles are essentially $\pm50^\circ$ on either side of opposition (Table \ref{tab.FG3obs}) and the sub-Earth latitudes on the surface of the body were at their maximum separation amongst the data considered here (Fig. \ref{fig.fg3therm}).

This raises a third and ultimately testable interpretation of the FG3 results. As shown in the bottom panel of Figure \ref{fig.fg3therm} the two epochs corresponded to two different viewing aspects: in 2009 the asteroid was at negative ecliptic latitude and thus observations preferentially viewed its northern hemisphere, whereas in 2011 the asteroid was at positive ecliptic latitude and thus observations preferentially accessed its southern hemisphere. The variability in optical and thermal properties could thus be suggestive of a heterogeneous surface. The ``top-shaped" morphology of FG3 with its pronounced equatorial ridge \citep{benner12,yu14} could serve to isolate the optical and thermal influence of the northern and southern hemispheres and thus accentuate surface heterogeneity even though the difference in viewing aspect was only $\sim20-30^\circ$ across the two epochs. If the surface of FG3 is heterogeneous, then we predict that the northern hemisphere would have optically neutral reflectance properties and a higher relative albedo, which would lower surface temperature and yield higher $\eta$. Conversely, the southern hemisphere would have optically red reflectance properties and a lower relative albedo, which would result in a higher relative surface temperature and correspondingly lower $\eta$. FG3 will be a strong optical target in 2022 and will be at negative ecliptic latitudes, thus providing a vantage of its northern hemisphere. Additional long term monitoring of this object may ultimately determine whether its measured variability is due to a heterogeneous surface \citep{deleon11,deleon13} or some other physical effect \citep{binzel12}. Spectral heterogeneity is not commonly seen on asteroids, particularly for km-scale objects like FG3, however near-IR observations of the 500-meter near-Earth object 101955 Bennu indicate variations in spectral slope that may be related to differences in regolith grain size across the surface \citep{binzel15}.


In light of the uncertainty surrounding our FG3 results it remains unclear whether physical properties like thermal inertia and pole orientation can be robustly derived from NEATM modeling of thermal emission in K-band. The highly non-uniform distribution of NEA obliquity, specifically a preference toward purely retrograde states \citep{laspina04}, would suggest that this technique should be well suited to detection of aspect-dependent thermal variability in the NEA population. Following on the work of \citet{delbo04}, new models to quantify the detectability of thermal asymmetry as a function of obliquity would provide new insight into the feasibility of such observational studies. Additionally, a larger sample size is needed to definitively determine whether NEATM modeling at near-IR wavelengths can robustly retrieve physical properties like pole orientation and thermal inertia.

It is clear that any future near-IR-based study of thermal inertia and pole orientation would require targets that span the widest possible range of phase angles on either side of opposition. On any given night\footnote{Based on a filtering of NEAs with the observability tool at http://asteroid.lowell.edu/upobjs.}, approximately 20\% of observable NEAs ($V<20$, solar elongation $>60^\circ$, galactic latitude $>20^\circ$) are at phase angles greater than $50^\circ$. Some of this uncertainty could be resolved by extending such phase dependent measurements to longer wavelengths, for instance photometry at L- and M-bands (3.5 and 4.5 $\mu$m respectively) or in the N- and Q-bands ($\sim10$ and 20 $\mu$m respectively) which are closer to the thermal emission peaks typical of NEOs. However, a systematic comparison of NEATM results based on data in different wavelength regimes is needed to more completely understand the limitations of this model. Extending measurements to longer wavelengths would have a primary benefit of enabling detection and modeling of thermal emission for any NEA, not just low albedo objects like QE2 and FG3. A targeted spectro-photometric survey to monitor phase-dependent thermal emission from select NEAs could produce an unprecedented sample of spin vector orientations. Such a sample would provide directs tests for models of NEA source regions in the Main Belt that depend on spin orientation and the subsequent direction of semi-major axis drift due to the Yarkovsky effect \citep[e.g.][]{bottke02}.

\ack

We are grateful to Eric Volquardsen for use of his NEATM thermal model code. This manuscript was significantly improved thanks to insightful reviews from Michael Mueller, Al Harris, and an anonymous referee. This work includes data obtained at NASA's IRTF located on Mauna Kea in Hawaii, which is operated by the University of Hawaii under contract NNH14CK55B with the National Aeronautics and Space Administration. Support for this project was provided through the National Science Foundation Astronomy and Astrophysics Postdoctoral Fellowship awarded to N.A.M and by NASA grant number NNX14AN82G issued through the Near-Earth Object Observations program. F.E.D. acknowledges support by the National Science Foundation under Grant 0907766 and by NASA under Grant NNX12AL26G and through Hubble Fellowship Grant HST-HF-51319.01-A awarded by the Space Telescope Science Institute, which is operated by the Association of Universities for Research in Astronomy, Inc., for NASA, under Contract NAS 5-26555. D.P. is thankful to the AXA research fund for support through their postdoctoral fellowship. E.S.H., R.J.V., abd Y.R.F. acknowledge partial support from National Science Foundation grant AST-1109855.

\label{lastpage}



\begin{thebibliography}{}

\bibitem[Benner et al.(2012)]{benner12} Benner, L.~A.~M. and 9 co-authors, 2012. Arecibo and Goldstone Radar Observations of Binary Near-Earth Asteroid and Marco Polo-R Mission Target (175706) 1996 FG3. American Astronomical Society, DPS meeting \#44, abstract 102.06. 

\bibitem[Binzel et al.(2001)]{binzel01} Binzel, R.~P., Harris, A.~W., Bus, S.~J., and Burbine, T.~H., 2001. Spectral Properties of Near-Earth Objects: Palomar and IRTF Results for 48 Objects Including Spacecraft Targets (9969) Braille and (10302) 1989 ML. Icarus 151, 139-149.

\bibitem[Binzel et al.(2012)]{binzel12} Binzel, R.~P., Polishook, D., DeMeo, F.~E., Emery, J.~P., Rivkin, A.~S., 2012. Marco Polo-R target asteroid (175706) 1996 FG3: Possible evidence for an annual thermal wave. Lunar Planet. Sci. 43. Abstract 1659.

\bibitem[Binzel et al.(2015)]{binzel15} Binzel, R.~P. and 16 co-authors, 2015. Spectral slope variations for OSIRIS-REx target Asteroid (101955) Bennu: Possible evidence for a fine-grained regolith equatorial ridge. Icarus 256, 22-29.

\bibitem[Bottke et al.(2002)]{bottke02} Bottke, W.~F., Morbidelli, A., Jedicke, R., Petit, J.~M., Levison, H.~F., Michel, P., Metcalfe, T., 2002. Debiased orbital and absolute magnitude distribution of near-Earth objects. Icarus 156, 399-433.

\bibitem[Bottke et al.(2006)]{bottke06} Bottke, W.~F., Vokroulick\'y, D., Rubincam, D.~P., Nesvorn\'y, D., 2006. The Yarkovsky andYORP Effects: Implications for asteroid dynamics. AREPS 34, 157-191.

\bibitem[Burbine(1998)]{burbine98} Burbine, T.~H., 1998. Could G-class asteroids be the parent bodies of CM chondrites? MAPS 33, 253-258.

\bibitem[Carry et al.(2010)]{carry10} Carry, B. and 10 co-authors, 2010. Physical properties of (2) Pallas. Icarus 205, 460-472.

\bibitem[Delbo et al.(2003)]{delbo03} Delbo, M., Harris, A.~W., Binzel, R.~P., Pravec, P., and Davies, J.~K., 2003. Keck observations of near-Earth asteroids in the thermal infrared. Icarus 166, 116-130.

\bibitem[Delbo(2004)]{delbo04} Delbo, M., 2004. The nature of near-Earth asteroids from the study of their thermal infrared emission. Doctoral thesis, Freie Universitat Berlin.

\bibitem[Delbo et al.(2007)]{delbo07} Delbo, M., dell'Oro, A., Harris, A.~W., Mottola, S., Mueller, M., 2007. Thermal inertia of near-Earth asteroids and implications for the magnitude of the Yarkovsky effect. Icarus 190, 236-249.

\bibitem[Delbo \& Tanga(2009)]{delbo09} Delbo, M. and Tanga, P., 2009. Thermal inertia of main belt asteroids smaller than 100 km from IRAS data. Planetary and Space Science 57, 259-265.

\bibitem[Delbo et al(2015)]{delbo15} Delbo, M., Mueller, M., Emery, J.~P., Rozitis, B.~E. and Capria, M.~T., 2015. Asteroid thermophysical modeling. In: Michel, P., DeMeo, F.~E., Bottke, W.~F. (Eds.), Asteroids IV. Univ. Arizona Press, Tucson, pp. 107-128.

\bibitem[de Leon et al.(2011)]{deleon11} de Leon, J., Mothe-Diniz, T., Licandro, J., Pinilla-Alonso, N., and Campins, H., 2011. New observations of asteroid (175706) 1996 FG3, primary target of the ESA Marco Polo-R mission. A\&A 530, L12.

\bibitem[de Leon et al.(2013)]{deleon13} de Leon, J., Lorenzi, V., Ali-Lagoa, V., Licandro, J., Pinilla-Alonso, N., and Campins, H., 2013. Additional spectra of asteroid 1996 FG3, backup target of the ESA MarcoPolo-R mission. A\&A 556, A33.

\bibitem[DeMeo et al.(2009)]{demeo09} DeMeo, F.~E., Binzel, R.~P., Slivan, S.~M., and Bus, S.~J., 2009. An extension of the Bus asteroid taxonomy into the near-infrared. Icarus 202, 160-180.

\bibitem[Farinella \& Vokrouhlick\'y(1999)]{farinella99} Farinella, P. and Vokrouhlick\'y, D., 1999. Semimajor axis mobility of asteroidal fragments. Science 283, 1507-1510.

\bibitem[Gehrels(1967)]{gehrels67} Gehrels, T., 1967. Minor Planets. I. The rotation of Vesta. AJ 72, 929-938.

\bibitem[Hansen(1977)]{hansen77} Hansen, O., 1977. On the prograde rotation of asteroids. Icarus 32, 458-460.

\bibitem[Harris(1998)]{harris98} Harris, A.~W., 1998. A thermal model for near-Earth asteroids. Icarus 131, 291-301.


\bibitem[Harris et al.(2007)]{harris07} Harris, A.~W., Mueller, M., Delbo, M., and Bus, S.~J., 2007. Physical characterization of the potentially hazardous high-albedo Asteroid (33342) 1998 WT24 from thermal-infrared observations. Icarus 188, 414-424.

\bibitem[Hicks et al.(2013)]{hicks13} Hicks, M., Lawrence, K., Chesley, S., Chesley, J., Rhoades, H., Elberhar, S., Carcione, A., Borlase, R., 2013. Palomar Spectroscopy of Near-Earth Asteroids 137199 (1999 KX4), 152756 (1999 JV3), 163249 (2002 GT), 163364 (2002 OD20), and 285263 (1998 QE2). The Astronomer's Telegram 5132, 1. 

\bibitem[Howell et al.(2015)]{howell15} Howell, E.~S. and 7 co-authors, 2015. SHERMAN: A Shape-based Thermophysical Model for Near-Earth Asteroids. American Astronomical Society, DPS meeting \#47, abstract 213.11. 


\bibitem[Jacobson \& Scheeres(2011)]{jacobson11} Jacobson, S.~A. and Scheeres, D.~J., 2011. Dynamics of rotationally fissioned asteroids: Source of observed small asteroid systems. Icarus 21, 161-178.

\bibitem[Johnson et al.(1983)]{johnson83} Johnson, P.~E., Kemp, J.~C., Lebofsky, M.~J., and Rike, G.~H., 1983. 10 $\mu m$ polarimetry of Ceres. Icarus 56, 381-392.

\bibitem[Kaasalainen et al.(2002)]{kaasalainen02} Kaasalainen, M., Torpa, J., Piironen, J., 2002. Models of twenty asteroids from photometric data. Icarus 159, 369-395.

\bibitem[La Spina et al.(2004)]{laspina04} La Spina, A., Paolicchi, P., Kryszczynska, A., and Pravec, P. 2004. Retrograde spins of near-Earth asteroids from the Yarkovsky effect. Nature 428, 400-401.


\bibitem[Lebofsky et al.(1986)]{lebofsky86} Lebofsky, L.~A. and 8 co-authors, 1986. A refined "standard" thermal model for asteroids based on observations of 1 Ceres and 2 Pallas. Icarus 68, 239-251.

\bibitem[Lowry et al.(2007)]{lowry07} Lowry, S.~C. and 10 co-authors, 2007. Direct detection of the asteroidal YORP effect. Science 316, 272-274.

\bibitem[Mainzer et al.(2011)]{mainzer11} Mainzer, A. and 36 co-authors, 2011. NEOWISE observations of near-Earth objects: Preliminary results. ApJ 743, 156.

\bibitem[Margot et al.(2002)]{margot02} Margot, J.~L. and 7 co-authors, 2002. Binary asteroids in the near-Earth object population. Science 296, 1445-1448.

\bibitem[Masiero et al.(2011)]{masiero11} Masiero, J.~R. and 17 co-authors, 2011. Main belt asteroids with WISE/NEOWISE. I. Preliminary albedos and diameters. ApJ 741, 68.

\bibitem[Matson(1971)]{matson71} Matson, D.~L., 1971. Infrared observations of asteroids. In {\it Physical Studies of Minor Planets} (T. Gehrels, Ed.), pp. 45-50. NASA SP-267.

\bibitem[Morrison(1976)]{morrison76} Morrison, D., 1976. The diameter and thermal inertia of 433 Eros. Icarus 28, 125-132.

\bibitem[Morrison(1977)]{morrison77} Morrison, D., 1977. Asteroid sizes and albedos. Icarus 31, 185-220.

\bibitem[Moskovitz et al.(2013)]{moskovitz13} Moskovitz, N.~A. and 11 co-authors, 2013. Rotational characterization of Hayabusa II target asteroid (162173) 1999 JU3. Icarus 224, 24-31.

\bibitem[Mueller et al.(2006)]{mueller06} Mueller, M., Harris, A.~W., Bus, S.~J., Hora, J.~L., Kassis, M., and Adams, J.~D., 2006. The size and albedo of Rosetta fly-by target 21 Lutetia from new IRTF measurements and thermal modeling. A\&A 447, 1153.

\bibitem[Mueller et al.(2011)]{mueller11} Mueller, M. and 16 co-authors, 2011. ExploreNEOs. III. Physical characterization of 65 potential spacecraft target asteroids. AJ 141, 109.

\bibitem[Muller et al.(2012)]{muller12} Muller, T.~G. and 8 co-authors, 2012. Physical properties of OSIRIS-REx target asteroid (101955) 1999 RQ36. A\&A 548, A36.

\bibitem[Nugent et al.(2012)]{nugent12} Nugent, C.~R., Margot, J.~L., Chesley, S.~R., Vokrouhlick\'y, D., 2012. Detection of semi major axis drifts in 54 near-Earth asteroids: New measurements of the Yarkovsky effect. AJ 144, 60.

\bibitem[Ostro et al(2002)]{ostro02} Ostro, S.~J., Hudson, R.~S., Benner, L.~A.~M., Giorgini, J.~D., Magri, C., Margot, J.~L., Nolan, M.~C., 2002. Asteroid radar astronomy. In: Bottke, W.~F., Cellino, A., Paolicchi, P., Binzel, R.~P. (Eds.), Asteroids III. Univ. Arizona Press, Tucson, pp. 151-168.

\bibitem[Paolicchi et al(2002)]{paolicchi02} Paolicchi, P., Burns, J.~A., and Weidenschilling, S.~J., 2002. Side Effects of Collisions: Spin Rate Changes, Tumbling Rotation States, and Binary Asteroids. In: Bottke, W.~F., Cellino, A., Paolicchi, P., Binzel, R.~P. (Eds.), Asteroids III. Univ. Arizona Press, Tucson, pp. 517-526.

\bibitem[Pieters(1983)]{pieters83} Pieters, C.~M., 1983. Strength of Mineral Absorption Featuresin the Transmitted Component of Near-Infrared Reflected Light: First Results From RELAB. J. Geophys. Res. 88, 9534-9544.

\bibitem[Polishook et al.(2012)]{polishook12} Polishook, D., Binzel, R.~P., Lockhart, M., DeMeo, F.~E., Golisch, W., Bus, S.~J., and Gulbis, A.~A.~S., 2012. Spectral and spin measurement of two small and fast-rotating near-Earth asteroids. Icarus 221, 1187-1189.

\bibitem[Pravec \& Harris(2000)]{pravec00} Pravec, P. and Harris, A.~W., 2000. Fast and slow rotation of asteroids. Icarus 148, 12-20.

\bibitem[Pravec et al.(2006)]{pravec06} Pravec, P. and 56 co-authors, 2006. Photometric survey of binary near-Earth asteroids. Icarus 181, 63-93.

\bibitem[Pravec et al.(2010)]{pravec10} Pravec, P. and 25 co-authors., 2010. Formation of asteroid pairs by rotational fission. Nature 466, 1085-1088.

\bibitem[Rayner et al.(2003)]{rayner03} Rayner, J.~T. and 7 co-authors, 2003. SpeX: A Medium-Resolution 0.8?5.5 Micron Spectrograph and Imager for the NASA Infrared Telescope Facility. PASP 115, 362-382.

\bibitem[Reddy et al.(2009)]{reddy09} Reddy, V., Emery, J.~P., Gaffey, M.~J., Bottke, W.~F., Cramer, A., and Kelley, M.~S. 2009. Composition of 298 Baptistina: Implications for the K/T impactor link. Meteor. Planet. Sci. 44, 1917-1927.

\bibitem[Reddy et al.(2012)]{reddy12} Reddy, V., Gaffey, M.~J., Abell, P.~A., and Hardersen, P.~S., 2012. Constraining albedo, diameter and composition of near-Earth asteroids via near-infrared spectroscopy. Icarus 219, 382-392.

\bibitem[Rivkin et al.(2005)]{rivkin05} Rivkin, A.~S., Binzel, R.~P., and Bus, S.~J., 2005. Constraining near-Earth object albedos using near-infrared spectroscopy. Icarus 175, 175-180.

\bibitem[Rivkin et al.(2012)]{rivkin12} Rivkin, A.~S. and 12 co-authors, 2012. The MarcoPolo-R target asteroid (175706) 1996 FG3: hydrated minerals and a variable spectral slope. European Planetary Science Congress, abstract 350. 

\bibitem[Rivkin et al.(2013)]{rivkin13} Rivkin, A.~S. and 8 co-authors, 2013. The NEO (175706) 1996 FG3 in the 2-4 $\mu m$ spectral region: Evidence for an aqueously altered surface. Icarus 223, 493-498.

\bibitem[Rubincam(2000)]{rubincam00} Rubincam, D.~P., 2000. Radiative spin-up and spin-down of small asteroids. Icarus 148, 2-11.

\bibitem[Scheeres(2007)]{scheeres07} Scheeres, D.~J., 2007. Rotational fision of contact binary asteroids. Icarus 189, 370-385.

\bibitem[Scheirich et al.(2015)]{scheirich15} Scheirich, P. and 25 co-authors, 2015. The binary near-Earth Asteroid (175706) 1996 FG3 ? An observational constraint on its orbital evolution. Icarus 245, 56-63.

\bibitem[Slivan(2002)]{slivan02} Slivan, S.~M., 2002. Spin vector allignment of Koronis family asteroids. Nature 419, 49-51.

\bibitem[Springmann et al.(2014)]{springmann14} Springmann, A. and 7 co-authors, 2014. Radar derived shape model of binary near-Earth asteroid (285263) 1998 QE2. Lunar Planet. Sci. 45. Abstract 1313.

\bibitem[Spencer et al.(1989)]{spencer89} Spencer, J.~R., Lebofsky, L.~A., and Sykes, M.~V., 1989. Systematic biases in radiometric diameter determinations. Icarus 78, 337-354.

\bibitem[Spencer(1990)]{spencer90} Spencer, J.~R., 1990. A rough-surface thermophysical model for airless planets. Icarus 83, 27-38.

\bibitem[Taylor et al.(2007)]{taylor07} Taylor, P.~A. and 11 co-authors, 2007. Spin rate of asteroid (54509) 2000 PH5 increasing due to the YORP effect. Science 316, 274-277.

\bibitem[Torppa et al.(2003)]{torppa03} Torppa, J. and 6 co-authors, 2003. Shapes and rotational properties of thirty asteroids from photometric data. Icarus 164, 346-383.

\bibitem[Trilling et al.(2010)]{trilling10} Trilling, D.~E. and 15 co-authors, 2010. ExploreNEOs. I. Description and first results from the warm Spitzer near-Earth object survey. AJ 140, 770-784.

\bibitem[Trilling et al.(2016)]{trilling16} Trilling, D.~E. and 14 co-authors, 2016. ExploreNEOs VI: Second data release and preliminary size distribution of near Earth objects. AJ, submitted.

\bibitem[Vokrouhlicky et al(2015)]{vok15} Vokrouhlicky, D., Bottke, W.~F, Chesley, S.~R., Scheeres, D.~J. and Statler, T.~S., 2015. The Yarkovsky and YORP Effects. In: Michel, P., DeMeo, F.~E., Bottke, W.~F. (Eds.), Asteroids IV. Univ. Arizona Press, Tucson, pp. 509-532.

\bibitem[Walsh et al.(2008)]{walsh08} Walsh, K.~J., Richardson, D.~C., Michel, P., 2008. Rotational breakup as the origin of small binary asteroids. Nature 454, 188-191.

\bibitem[Walsh et al.(2012)]{walsh12} Walsh, K.~J., Delbo, M., Mueller, M., Binzel, R.~P., and DeMeo, F.~E., 2012. Physical characterization and origin of binary near-Earth asteroid (175706) 1996 FG3. ApJ 748, 104.

\bibitem[Warner et al.(2009)]{warner09} Warner, B.~D., Harris, A.~W., and Pravec, P., 2009. The asteroid lightcurve database. Icarus 202, 134-146.

\bibitem[Wolters et al.(2008)]{wolters08} Wolters, S.~D., Green, S.~F., McBride, N., and Davies J.~K., 2008. Thermal infrared and optical observations of four near-Earth asteroids. Icarus 193, 535-552.

\bibitem[Wolters \& Green(2009)]{wolters09} Wolters, S.~D., and Green, S.~F., 2009. Investigation of systematic bias in radiometric diameter determination of near-Earth asteroids: the night emission simulated thermal model (NESTM). MNRAS 400, 204-218.

\bibitem[Wolters et al.(2011)]{wolters11} Wolters, S.~D., and 7 co-authors, 2011. Physical characterization of low delta-V asteroid (175706) 1996 FG3. MNRAS 418, 1246-1257.

\bibitem[Yu et al.(2014)]{yu14} Yu, L., Ji, J., and Wang S., 2014. Shape, thermal and surface properties determination of a candidate spacecraft target asteroid (175706) 1996 FG3. MNRAS 439, 3357-3370.

\bibitem[Zuber et al.(2000)]{zuber00} Zuber, M.~T. and 11 co-authors, 2000. The Shape of 433 Eros from the NEAR-Shoemaker Laser Rangefinder. Science 289, 2097-2101.

\end{thebibliography}




\clearpage	



\end{document}